\documentclass[aps,prd,epsecnum,nofootinbib,11pt,preprintnumbers]{revtex4}
\usepackage{amsmath,amsthm,amssymb}
\usepackage[dvipdfm]{graphicx}
\begin{document}
\title{Heterotic Solutions with $G_{2}$ and Spin(7) Structures}

\author{
Kazuki Hinoue\footnote{E-mail: hinoue@sci.osaka-cu.ac.jp},
Yukinori Yasui\footnote{E-mail: yasui@sci.osaka-cu.ac.jp}
}
\affiliation{
Department of Mathematics and Physics, Graduate School of Science, Osaka City University, Osaka, Japan
}


\begin{abstract}
We study supersymmetric solutions in 7- and 8-dimensional Abelian heterotic supergravity theories.
In dimension 7, the solutions are described by $G_{2}$ with torsion equations.
When a $G_{2}$ manifold has principal orbits $S^{3} \times S^{3}$, the equations are reduced to ordinary differential equations with four radial functions.
For these equations we obtain explicit ALC metrics with $S^{3}$-bolt and $T^{1,1}$-bolt singularities. 
In dimension 8, we study supersymmetric solutions to $Spin(7)$ with torsion equations associated with 3-Sasakian manifolds by using a similar method to the case $G_{2}$.
\end{abstract}


\preprint{OCU-PHYS 414}
\maketitle
\section{Introduction}
The supersymmetric equations of supergravity theory are expressed as Killing spinor equations.
Specifically, the geometry of a Neveu--Schwarz(NS-NS) sector in type II supergravity theory and an NS sector in heterotic supergravity theory can be described in terms of $G$-structures.
The use of $G$-structures rehashes the supersymmetric equations as equations rewritten by means of differential forms. 
A $G$-structure has provided us with a technique for constructing supersymmetric solutions for these theories\cite{GMPW}\cite{BGGG}.

In this paper, we study supersymmetric solutions in 7- and 8-dimensional Abelian heterotic supergravity theories, 
where a $G_{2}$-structure on $7$-dimensional manifolds and a $Spin(7)$-structure on $8$-dimensional manifolds play an important role.
The space of intrinsic torsion can be decomposed into the irreducible components associated with representations of $G_{2}$ or $Spin(7)$.
The classification of $G_{2}$- or $Spin(7)$-structures is given by all possible combinations of these irreducible components.
The different classes of these structures are characterized via the differential equations of their $G$-invariant forms.
The equations of the specific classes associated with supergravity theories give the necessary and sufficient conditions for supersymmetry preservation in the theories\cite{GMW}\cite{GKMW}.
If there exist $G$-invariant forms satisfying the conditions, then they solve the equations of motion of common sectors in type II and heterotic supergravity theory\cite{II}\cite{I}.

When $G_{2}$-invariant 3-form $\Omega$ satisfies the condition
\begin{equation}\label{FG}
d\Omega = 0 \;,\; d*\Omega = 0 \,,
\end{equation}
the corresponding $G_{2}$ metric is a Ricci-flat metric with $G_{2}$ holonomy\cite{Fernandez-Gray}, 
which is obtained by solving the first order equations (\ref{FG}).
For the principal orbits $\textrm{\boldmath $CP$}(3)$ or $S^{3} \times S^{3}$, the $G_{2}$ metrics have been studied\cite{BGGG}\cite{BS}\cite{GPP}\cite{CGLP1}\cite{CGLP2}. Imposed on
a cohomogeneity one ansatz, equations (\ref{FG}) reduce to ordinary differential equations.
These equations were explicitly solved and asymptotically conical(AC) metrics\cite{BS}\cite{GPP} and asymptotically locally conical(ALC) metrics\cite{BGGG}\cite{CGLP1}\cite{CGLP2} were constructed.
Similarly, the $Spin(7)$-structure is given by the condition
\begin{equation}
d\Psi = 0 \,,
\end{equation}
where $\Psi$ is a $Spin(7)$-invariant 4-form which satisfies $*\Psi = \Psi$ \cite{F}.
The corresponding metrics with $Spin(7)$ holonomy are obtained in a similar method to the $G_{2}$ metrics.
For $Sp(2)/Sp(1)$ or $SU(3)/U(1)$ principal orbits, the $Spin(7)$ metrics have been studied\cite{CGLP1}\cite{CGLP2}\cite{CGLP3}\cite{CGLP4}\cite{KY1}\cite{KY2}.

By solving the defining equations for the $G_{2}$-structure in the class $W_{7} \oplus W_{27}$, supersymmetric solutions
of the form $\textrm{\boldmath $R$}^{1,2} \times M_{7}$ in type II supergravity theory were constructed\cite{GMPW},
where the flux $H$ and the dilaton $\Phi$ were identified with 3-form torsion and Lee form, respectively.
Similar to these solutions, we will solve the defining equations of the $G_{2}$ or $Spin(7)$-structure in the associated class with Abelian heterotic supergravity theory to obtain supersymmetric solutions.
Unlike type II supergravity theory, Abelian heterotic supergravity theory additionally has a $U(1)$ gauge field.
However, the field strength of the $U(1)$ gauge field is determined algebraically and thus we will use a similar method to derive the supersymmetric solutions in Abelian heterotic theory.
This will lead us to solutions in heterotic supergravity theory by recovering $\alpha'$ collection and using modified standard embedding\cite{MS}.
Solutions that would be helpful in studying heterotic gauge fivebranes wrapping associative three cycles are expected\cite{GMPW}.
Thus, we will construct the supersymmetric solutions in Abelian heterotic theory.

The remainder of this paper is organized as follows.
We begin section 2 with a short explanation about Abelian heterotic supergravity theory.
We also review a $G_{2}$-structure associated with the heterotic supergravity theory. 
In section 3, we study the $G_{2}$ with torsion equations on cohomogeneity one manifold $\textrm{\boldmath $R$}_{+} \times S^{3} \times S^{3}$ and derive the supersymmetric 
solutions for the resulting ordinary differential equations.
In section 4, we review a $Spin(7)$-structure associated with the heterotic supergravity theory and
derive the $Spin(7)$ with torsion equations on 8-dimensional manifold $\textrm{\boldmath $R$}_{+} \times P$, where $P$ is a 3-Sasakian manifold.
The supersymmetric solutions are briefly discussed by using a similar method to the $G_2$ case.
The final section 5 is devoted to conclusion.
\section{$G_{2}$-structure associated with supergravity}
 In this section, we review a $G_{2}$-structure related with 7-dimensional Abelian heterotic supergravity theory.
We first explain Abelian heterotic supergravity theory shortly.
Next, we introduce the $G_{2}$-structure and derive the fundamental equations describing supersymmetric equations.
\subsection{Abelian heterotic supergravity}
 Let us consider the low energy effective theory with an $\alpha'$ expansion of heterotic string theory.
In this theory restricting the gauge group to $U(1)$ and setting $\alpha'=1$, one obtains 10-dimensional Abelian heterotic supergravity theory.
This system consists of a metric $g$, dilaton $\Phi$, $U(1)$ gauge field $A$ and $B$ field.
Then the string frame Lagrangian is given by
\begin{equation}
\mathcal{L} = e^{-\Phi}\left(R*1 + *d\Phi \wedge d\Phi - *F \wedge F - \frac{1}{2}*H \wedge H\right) \,, \label{abhL}
\end{equation}
where $R$ represents a scalar curvature and $F = dA$. 
The 3-form flux $H$ is given by $H = dB + A \wedge dA$, which satisfies the Bianchi identity
\begin{equation}
dH = F \wedge F \text{.} \label{abhBianchi}
\end{equation}
In addition, $*$ represents the Hodge dual operator associated with the metric $g$.
The equations of motion are written as
\begin{eqnarray}
R_{\mu\nu} & = & - \nabla _{\mu}\nabla _{\nu}\Phi + F_{\mu}\,^{\rho}F_{\nu\rho} + \frac{1}{4}H_{\mu}\,^{\rho\sigma}H_{\nu\rho\sigma} \,, \label{eqm1} \\
\nabla ^{2}e^{-\Phi} & = & \frac{1}{2}e^{-\Phi}F_{\mu\nu}F^{\mu\nu} + \frac{1}{6}e^{-\Phi}H_{\mu\nu\rho}H^{\mu\nu\rho} \,, \label{eqm2} \\
d(e^{-\Phi}*F) & = & - e^{-\Phi}*H \wedge F \,, \label{eqm3} \\
d(e^{-\Phi}*H) & = & 0 \label{eqm4}
\end{eqnarray}
and the corresponding supersymmetric equations are
\begin{eqnarray}
\nabla ^{T}_{\mu}\chi  \equiv  \left(\nabla_{\mu} + \frac{1}{4 \cdot 2!}H_{\mu\nu\rho}\gamma^{\nu\rho}\right)\chi = 0 \,, \label{SUSY1} \\
(\partial _{\mu}\Phi)\gamma^{\mu}\chi + \frac{1}{3!}H_{\mu\nu\rho}\gamma^{\mu\nu\rho}\chi = 0 \,, \label{SUSY2} \\
F_{\mu\nu}\gamma^{\mu\nu}\chi  =  0 \,, \label{SUSY3} 
\end{eqnarray}
where $\chi$ denotes a Killing spinor.
These equations (\ref{SUSY1}), (\ref{SUSY2}) and (\ref{SUSY3}) are derived from the supersymmetry variation of gravitino, dilatino and gaugino, respectively.

It is known that if the fields satisfy the supersymmetric equations and Bianchi identity, they automatically solve the equations of motion \cite{II}\cite{I}.
In this paper, we shall construct supersymmetric solutions in the heterotic theory.
For this purpose, we consider 7- and 8-dimensional manifolds $M_{n}$ admitting $G_{2}$ and $Spin(7)$ structures
\footnote{For 6-dimensional case, supersymmetric solutions of Abelian heterotic theory were constructed in the view of supergravity solution generating method \cite{MS}.}.
The corresponding 10-dimensional spacetimes are assumed to be of the form $\textrm{\boldmath $R$}^{1,9-n} \times M_{n}~(n=7,8)$ where the fields are non-trivial only on $M_{n}$.
\subsection{$G_{2}$-structure}
 A $G_{2}$-structure over $M_{7}$ is a principal subbundle with fiber $G_{2}$ of the frame bundle $\mathcal{F}(M_{7})$ \cite{Joyce}.
There is a one to one correspondence between $G_{2}$ structures and $G_{2}$ invariant 3-forms $\Omega$.
The standard form of $\Omega$ is given by
\begin{equation}
\Omega = e^{123} + e^{147} + e^{165} + e^{246} + e^{257} + e^{354} + e^{367} \,, \label{ff3G2}
\end{equation}
where $e^{\mu}$ $(\mu = 1,\cdots,7)$ is an orthonormal frame of the metric, $g = \sum_{\mu = 1}^{7}e^{\mu} \otimes e^{\mu}$ and $e^{\mu\nu\rho} = e^{\mu} \wedge e^{\nu} \wedge e^{\rho}$.

We consider the $G_{2}$-structure $\Omega$ associated with heterotic supergravity. 
According to \cite{FI}, we require
\begin{eqnarray}
d*\Omega & = & \Theta \wedge *\Omega \,, \label{G2defeq1}
\end{eqnarray}
where $\Theta$ is the Lee form defined by 
\begin{equation}
\Theta = - \frac{1}{3}*(*d\Omega \wedge \Omega) \,. \label{Lee}
\end{equation}
Then we have a unique connection $\nabla^{T}$ preserving the $G_{2}$-structure with 3-form torsion\cite{II}\cite{FI}\cite{FIUV}
\begin{equation}
T = \frac{1}{6}*(d\Omega \wedge \Omega)\Omega - *\left(d\Omega - \Theta \wedge \Omega\right) \,, \label{G2torsion}
\end{equation}
i.e. $\nabla^{T}\Omega=\nabla^T g = 0$.
This structure belongs to the class $W_{1} \oplus W_{7} \oplus W_{27}$ in the classification by Fern$\acute{\text{a}}$ndez and Gray \cite{Fernandez-Gray}(see also \cite{S.Karigiannis}).
It was shown in the papers \cite{FI},\cite{FIUV} that there exists a non-trivial solution to both dilatino and gravitino Killing spinor equations in dimension 7 if and only if there exists a $G_{2}$-structure satisfying (\ref{G2defeq1}) and 
\begin{eqnarray}
d\Omega \wedge \Omega & = & 0 \label{G2defeq2} 
\end{eqnarray}
together with an exact Lee form $\Theta$.
Here (\ref{G2defeq2}) yields that this $G_{2}$-structure is in the class $W_{7} \oplus W_{27}$.
If $\Theta = 0$ and $T = 0$ one can find $d\Omega = 0$ and $d*\Omega = 0$, which are equivalent to $\nabla \Omega = 0$ and the corresponding metric is a Ricci-flat metric with $G_{2}$ holonomy.
Now, we have the following identification
\begin{eqnarray}
& & H = T \label{identification1} \,, \\
& & d\Phi = \Theta \,. \label{identification2}
\end{eqnarray}
Thus we obtain
\begin{eqnarray}
H = - e^{\Phi}*d(e^{-\Phi}\Omega) \,,~~\, d\Phi = - \frac{1}{3}*(*d\Omega \wedge \Omega) \,, 
\end{eqnarray}
which means vanishing of supersymmetry variations for gravitino and dilatino, respectively.
In addition, the supersymmetry variation for gaugino requires the generalized self-dual condition on the field strength $F$,
\begin{equation}
*(\Omega \wedge F) = F \,. \label{gs}
\end{equation}
In general, 2-forms on $M_{7}$ are decomposed into $G_{2}$ irreducible representations under action of $G_{2}$, 
\begin{eqnarray}
\wedge ^{2}_{7} & = & \{\alpha \in \wedge ^{2} \,:\, *(\Omega \wedge \alpha) = - 2\alpha\} \,, \\
\wedge ^{2}_{14} & = & \{\alpha \in \wedge ^{2} \,:\, *(\Omega \wedge \alpha) = \alpha\} \,,
\end{eqnarray}
where subscripts $7,14$ of $\wedge^2 $ are dimensions of the representation.
Thus the field strength $F$ in (\ref{gs}) is in $\wedge ^{2}_{14}$.
The Bianchi identity (\ref{abhBianchi}) gives a relation between $T$ and $F$, which leads to a strong restriction for the field strength $F$.

In summary, supersymmetric equations associated with Abelian heterotic supergravity theory are given by
\begin{eqnarray}
& &e^{\Phi}d(e^{-\Phi}*\Omega)= 0 \label{defining1} \,, \\
& &d\Omega \wedge \Omega = 0 \label{defining2} \,, \\
& &H = - e^{\Phi}*d(e^{-\Phi}\Omega) \label{gravitino} \,, \\
& &d\Phi= - \frac{1}{3}*(*d\Omega \wedge \Omega) \label{dilatino} \,, \\
& &*(\Omega \wedge F)= F \label{gaugino} \,, \\
& &dH  =  F \wedge F \,. \label{Bianchi}
\end{eqnarray}
The first equation is derived from (\ref{G2defeq1}) and (\ref{identification2}) easily.
In the following we call these equations $G_{2}$ with torsion equations or simply $G_2T$ equations.
Thus it is enough  to solve $G_{2}T$ equations so as to construct supersymmetric solutions.\\

\section{$G_2$ solutions in 7-dimensional Abelian heterotic supergravity}
 In this section, we study $G_{2}T$ equations under cohomogeneity one ansatz.
Then these equations are reduced to first order non-linear ordinary differential equations.
We solve reduced equations numerically and also try to construct analytically solutions for these equations.
\subsection{Cohomogeneity one ansatz}
The $G_{2}T$ equations can be reduced to ordinary differential equations under cohomogeneity one ansatz.
A $G$-manifold is called cohomogeneity one manifold if the principal orbits are hypersurfaces.
It is known that cohomogeneity one manifolds admitting $G_{2}$-structures are classified into seven types \cite{CS}. 
In this paper we will consider the case when the principal orbits are $S^{3} \times S^{3}$.
Then the manifold is locally $\textrm{\boldmath $R$}_{+} \times S^{3} \times S^{3}$.
Similar analysis can be done for the other orbits.

We consider the following metric with six radial functions $a_{i}(t),b_{i}(t) \; (i = 1,2,3)$
\begin{equation}
g = dt^{2} + \sum_{i=1}^{3}a_{i}(t)^{2}(\sigma_{i} - \Sigma_{i})^{2} + \sum_{i=1}^{3}b_{i}(t)^{2}(\sigma_{i} + \Sigma_{i})^{2} \,, \label{6fmetric}
\end{equation}
where $\sigma_{i}$ and $\Sigma_{i}$ are left invariant 1-forms on two SU(2) group manifolds which satisfy the relations
\begin{eqnarray}
d\sigma_{1} & = & - \sigma_{2} \wedge \sigma_{3} \;,\; d\sigma_{2} = - \sigma_{3} \wedge \sigma_{1} \;,\; d\sigma_{3} = - \sigma_{1} \wedge \sigma_{2} \,, \nonumber \\
d\Sigma_{1} & = & - \Sigma_{2} \wedge \Sigma_{3} \;,\; d\Sigma_{2} = - \Sigma_{3} \wedge \Sigma_{1} \;,\; d\Sigma_{3} = - \Sigma_{1} \wedge \Sigma_{2} \,.
\end{eqnarray}
It is convenient to introduce an orthonormal frame $e^{\mu}$ $(\mu = 1, \cdots, 7)$ defined by 
\begin{eqnarray}
e^{1} & = & a_{1}(t)(\sigma_{1} - \Sigma_{1}) \;,\; e^{2} = a_{2}(t)(\sigma_{2} - \Sigma_{2}) \;,\; e^{3} = a_{3}(t)(\sigma_{3} - \Sigma_{3}) \,, \nonumber \\
e^{4} & = & b_{1}(t)(\sigma_{1} + \Sigma_{1}) \;,\; e^{5} = b_{2}(t)(\sigma_{2} + \Sigma_{2}) \;,\; e^{6} = b_{3}(t)(\sigma_{3} + \Sigma_{3}) \;,\; e^{7} = dt \,. 
\end{eqnarray}
Then the $G_{2}$-structure $\Omega$ takes the form (\ref{ff3G2}), which automatically satisfy (\ref{defining2}).
The dilaton $\Phi$ is explicitly calculated from (\ref{dilatino}), 
\begin{eqnarray}
\frac{d\Phi}{dt} & = & - \frac{1}{3}\left( - \frac{2}{a_{1}}\frac{da_{1}}{dt} - \frac{2}{a_{2}}\frac{da_{2}}{dt} - \frac{2}{a_{3}}\frac{da_{3}}{dt} - \frac{2}{b_{1}}\frac{db_{1}}{dt} - \frac{2}{b_{2}}\frac{db_{2}}{dt} - \frac{2}{b_{3}}\frac{db_{3}}{dt} \right. \nonumber\\
                 &   &   + \frac{a_{1}}{2a_{2}b_{3}} + \frac{a_{1}}{2a_{3}b_{2}} + \frac{a_{2}}{2a_{3}b_{1}} + \frac{a_{2}}{2a_{1}b_{3}} + \frac{a_{3}}{2a_{1}b_{2}} + \frac{a_{3}}{2a_{2}b_{1}} \nonumber \\ 
                 &   &    \left. + \frac{b_{1}}{2a_{2}a_{3}} - \frac{b_{1}}{2b_{2}b_{3}} + \frac{b_{2}}{2a_{1}a_{3}} - \frac{b_{2}}{2b_{1}b_{3}} + \frac{b_{3}}{2a_{1}a_{2}} - \frac{b_{3}}{2b_{1}b_{2}}\right) \,. \label{dPhidt}
\end{eqnarray}
The 3-form flux $H = H_{126}e^{126} + H_{456}e^{456} + H_{135}e^{135} + H_{234}e^{234}$ is also calculated from (\ref{gravitino}),
\begin{eqnarray}
H_{126} & = & \frac{1}{a_{3}}\frac{da_{3}}{dt} + \frac{1}{b_{1}}\frac{db_{1}}{dt} + \frac{1}{b_{2}}\frac{db_{2}}{dt} - \frac{a_{1}}{2a_{3}b_{2}} - \frac{a_{2}}{2a_{3}b_{1}} + \frac{b_{3}}{2b_{1}b_{2}} - \frac{d\Phi}{dt} \,, \nonumber \\
H_{456} & = & - \frac{1}{a_{1}}\frac{da_{1}}{dt} - \frac{1}{a_{2}}\frac{da_{2}}{dt} - \frac{1}{a_{3}}\frac{da_{3}}{dt} + \frac{b_{1}}{2a_{2}a_{3}} + \frac{b_{2}}{2a_{1}a_{3}} + \frac{b_{3}}{2a_{1}a_{2}} + \frac{d\Phi}{dt}  \,, \nonumber \\
H_{135} & = & - \frac{1}{a_{2}}\frac{da_{2}}{dt} - \frac{1}{b_{1}}\frac{db_{1}}{dt} - \frac{1}{b_{3}}\frac{db_{3}}{dt} + \frac{a_{1}}{2a_{2}b_{3}} - \frac{b_{2}}{2b_{1}b_{3}} + \frac{a_{3}}{2a_{2}b_{1}} + \frac{d\Phi}{dt} \,, \nonumber \\
H_{234} & = & \frac{1}{a_{1}}\frac{da_{1}}{dt} + \frac{1}{b_{2}}\frac{db_{2}}{dt} + \frac{1}{b_{3}}\frac{db_{3}}{dt} + \frac{b_{1}}{2b_{2}b_{3}} - \frac{a_{2}}{2a_{1}b_{3}} - \frac{a_{3}}{2a_{1}b_{2}} - \frac{d\Phi}{dt} \,, \label{difgravitino}
\end{eqnarray}
which clearly depend on only $t$.

One may think that the ansatz (\ref{6fmetric}) is artificial.
However the ansatz (\ref{6fmetric}) is reasonable from the view of a half-flat structure.
A $G_{2}$ structure and 7-dimensional metric on $\textrm{\boldmath $R$}_{+} \times S^{3} \times S^{3}$ are given by a one-parameter family of a half-flat structure on $S^{3} \times S^{3}$ \cite{T.B.M and S}.
A half-flat structure is one of $SU(3)$-structures, which is characterized by a pair $(\kappa,\gamma)$, where $\kappa$ is a symplectic form on $S^{3} \times S^{3}$
and $\gamma$ is a real 3-form whose stabiliser is isomorphic to $SL(3;\textrm{\boldmath $C$})$ and satisfies a condition $\gamma \wedge \kappa = 0$.
If $(\kappa,\gamma)$ satisfies the defining equation
\begin{equation}
d(\kappa \wedge \kappa) = 0 \;,\; d\gamma = 0 \text{,}
\end{equation}
then the $SU(3)$-structure is called a half-flat structure\cite{LL}.
Indeed the following half-flat pair $(\kappa,\gamma)$,
\begin{eqnarray}
\kappa & = & p_{1}\sigma_{1} \wedge \Sigma_{1} + p_{2}\sigma_{2} \wedge \Sigma_{2} + p_{3}\sigma_{3} \wedge \Sigma_{3} \,, \\
\gamma & = & q_{1}\sigma_{1} \wedge \sigma_{2} \wedge \sigma_{3} + q_{2}\Sigma_{1} \wedge \Sigma_{2} \wedge \Sigma_{3} + q_{3}(\sigma_{2} \wedge \sigma_{3} \wedge \Sigma_{1} - \sigma_{1} \wedge \Sigma_{2} \wedge \Sigma_{3}) \nonumber \\
       &   & + q_{4}(\sigma_{3} \wedge \sigma_{1} \wedge \Sigma_{1} - \sigma_{2} \wedge \Sigma_{3} \wedge \Sigma_{1}) + q_{5}(\sigma_{1} \wedge \sigma_{2} \wedge \Sigma_{3} - \sigma_{3} \wedge \Sigma_{1} \wedge \Sigma_{2})
\end{eqnarray}
leads to the six dimensional part of the metric (\ref{6fmetric}) under certain parameters $p_{i}$ and $q_{j}$.

The field strength $F$ is restricted by (\ref{gaugino}) and (\ref{Bianchi}).
Actually, we find $F$ is classified into seven types, 
\begin{eqnarray}
& & F_{12}e^{12} + F_{45}e^{45} + F_{67}e^{67} \;,\; F_{13}e^{13} + F_{46}e^{46} + F_{57}e^{57} \,, \nonumber \\
& & F_{23}e^{23} + F_{47}e^{47} + F_{56}e^{56} \;,\; F_{14}e^{14} + F_{25}e^{25} + F_{36}e^{36} \,, \nonumber \\
& & F_{15}e^{15} + F_{24}e^{24} + F_{37}e^{37} \;,\; F_{16}e^{16} + F_{27}e^{27} + F_{34}e^{34} \,, \nonumber \\
& & F_{17}e^{17} + F_{26}e^{26} + F_{35}e^{35} \,,
\end{eqnarray}
together with generalized self-dual relations
\begin{eqnarray}
F_{13} - F_{46} & = & F_{57} \;,\; F_{56} - F_{23} = F_{47} \;,\; F_{45} - F_{12} = F_{67} \,, \nonumber \\
F_{36} - F_{14} & = & F_{25} \;,\; F_{24} - F_{15} = F_{37} \;,\; F_{16} - F_{27} = F_{34} \;,\; F_{35} - F_{26} = F_{17} \,. 
\end{eqnarray}
Let us restrict ourselves to the case
\begin{equation}
F = F_{12}e^{12} + F_{45}e^{45} + F_{67}e^{67} \,. \label{F}
\end{equation}
Without loss of generality this case can be selected because other cases are obtained by using discrete symmetries of $G_{2}$T equations
\footnote{The field strength $F = F_{14}e^{14} + F_{25}e^{25} + F_{36}e^{36}$ is exceptional. In this case, $G_{2}T$ equations lead to the solution $F = 0$, which was studied in the paper\cite{GMPW}.}.
For exmaple, $G_{2}T$ equations are invariant under the following transformation, 
\begin{eqnarray}
& & F_{12} \longrightarrow F_{23} \;,\; F_{45} \longrightarrow F_{56} \;,\; F_{67} \longrightarrow  F_{47} \,, \nonumber \\
& & a_{1} \longrightarrow a_{2} \;,\; a_{2} \longrightarrow a_{3} \;,\; a_{3} \longrightarrow a_{1} \;,\; b_{1} \longrightarrow b_{2} \;,\; b_{2} \longrightarrow b_{3} \;,\; b_{3} \longrightarrow b_{1} \,, \nonumber \\
& & H_{126} \longrightarrow H_{234} \;,\; H_{135} \longrightarrow - H_{126} \;,\; H_{456} \longrightarrow H_{456} \;,\; H_{234} \longrightarrow - H_{135} \,. 
\end{eqnarray}

We have already written down (\ref{defining2}), (\ref{gravitino}), (\ref{dilatino}) and (\ref{gaugino}) under the cohomogeneity one ansatz.
The remaining $G_{2}T$ equation we should solve are (\ref{defining1}) and (\ref{Bianchi}).
The former is given by
\begin{eqnarray}
\frac{1}{a_{1}}\frac{da_{1}}{dt} + \frac{1}{a_{2}}\frac{da_{2}}{dt} + \frac{1}{b_{1}}\frac{db_{1}}{dt} + \frac{1}{b_{2}}\frac{db_{2}}{dt} - \frac{b_{3}}{2a_{1}a_{2}} - \frac{a_{3}}{2a_{1}b_{2}} - \frac{a_{3}}{2a_{2}b_{1}} + \frac{b_{3}}{2b_{1}b_{2}} - \frac{d\Phi}{dt} & = & 0 \,, \nonumber \\
\frac{1}{a_{1}}\frac{da_{1}}{dt} + \frac{1}{a_{3}}\frac{da_{3}}{dt} + \frac{1}{b_{1}}\frac{db_{1}}{dt} + \frac{1}{b_{3}}\frac{db_{3}}{dt} - \frac{b_{2}}{2a_{1}a_{3}} - \frac{a_{2}}{2a_{1}b_{3}} + \frac{b_{2}}{2b_{1}b_{3}} - \frac{a_{2}}{2a_{3}b_{1}} - \frac{d\Phi}{dt} & = & 0 \,, \nonumber \\
\frac{1}{a_{2}}\frac{da_{2}}{dt} + \frac{1}{a_{3}}\frac{da_{3}}{dt} + \frac{1}{b_{2}}\frac{db_{2}}{dt} + \frac{1}{b_{3}}\frac{db_{3}}{dt} - \frac{b_{1}}{2a_{2}a_{3}} + \frac{b_{1}}{2b_{2}b_{3}} - \frac{a_{1}}{2a_{2}b_{3}} - \frac{a_{1}}{2a_{3}b_{2}} - \frac{d\Phi}{dt} & = & 0 \,, 
\label{dPhi}
\end{eqnarray}
and the latter is given by
\begin{eqnarray}
\frac{dH_{456}}{dt} + H_{456}\left(\frac{1}{b_{1}}\frac{db_{1}}{dt} + \frac{1}{b_{2}}\frac{db_{2}}{dt} + \frac{1}{b_{3}}\frac{db_{3}}{dt}\right) & = & - 2F_{45}F_{67} \,, \nonumber \\
\frac{dH_{126}}{dt} + H_{126}\left(\frac{1}{a_{1}}\frac{da_{1}}{dt} + \frac{1}{a_{2}}\frac{da_{2}}{dt} + \frac{1}{b_{3}}\frac{db_{3}}{dt}\right) & = & - 2F_{12}F_{67} \,, \nonumber\\
\frac{dH_{135}}{dt} + H_{135}\left(\frac{1}{a_{1}}\frac{da_{1}}{dt} + \frac{1}{a_{3}}\frac{da_{3}}{dt} + \frac{1}{b_{2}}\frac{db_{2}}{dt}\right) & = & 0 \,, \nonumber \\
\frac{dH_{234}}{dt} + H_{234}\left(\frac{1}{a_{2}}\frac{da_{2}}{dt} + \frac{1}{a_{3}}\frac{da_{3}}{dt} + \frac{1}{b_{1}}\frac{db_{1}}{dt}\right) & = & 0 \,, \label{difBianchi} \\
H_{126}\frac{a_{1}}{2a_{3}b_{2}} - H_{456}\frac{b_{1}}{2a_{2}a_{3}} - H_{135}\frac{a_{1}}{2a_{2}b_{3}} - H_{234}\frac{b_{1}}{2b_{2}b_{3}} & = & 0 \,, \nonumber \\
H_{126}\frac{a_{2}}{2a_{3}b_{1}} - H_{456}\frac{b_{2}}{2a_{1}a_{3}} + H_{135}\frac{b_{2}}{2b_{1}b_{3}} + H_{234}\frac{a_{2}}{2a_{1}b_{3}} & = & 0 \,, \nonumber \\
H_{126}\frac{b_{3}}{2b_{1}b_{2}} + H_{456}\frac{b_{3}}{2a_{1}a_{2}} + H_{135}\frac{a_{3}}{2a_{2}b_{1}} - H_{234}\frac{a_{3}}{2a_{1}b_{2}} & = & - 2F_{12}F_{45} \,, \nonumber
\end{eqnarray}
which consist of differential equations and algebraic equations.
The algebraic equations yields the following relations,
\begin{eqnarray}
H_{456} & = & \frac{a_{1}a_{2}}{b_{1}b_{2}}H_{126} \;,\; H_{234} = - \frac{a_{1}b_{2}}{a_{2}b_{1}}H_{135} \,, \nonumber \\
F_{12} & = & \pm \sqrt{- \frac{1}{2}\left(\frac{b_{3}}{a_{1}a_{2}}H_{126} + \frac{a_{3}b_{2}}{a_{1}a_{2}\,^{2}}H_{135}\right)} \,, \nonumber \\
F_{45} & = & \frac{a_{1}a_{2}}{b_{1}b_{2}}F_{12} \;,\; F_{67} = F_{45} - F_{12}  \label{alg equations}
\end{eqnarray}
and $a_{i}$, $b_{i}$, $H_{126}$ and $H_{135}$ are determined by the differential equations (\ref{difgravitino}), (\ref{dPhi}) and (\ref{difBianchi}) :
\begin{equation}
\frac{da_{1}}{dt} = \frac{- a_{1}\,^{2} + a_{2}\,^{2} + b_{3}\,^{2}}{4a_{2}b_{3}} + \frac{- a_{1}\,^{2} + a_{3}\,^{2} + b_{2}\,^{2}}{4a_{3}b_{2}} - \frac{1}{2}a_{1}H_{126} + \frac{1}{2}a_{1}H_{135} \,, \label{ODE1}
\end{equation}
\begin{equation}
\frac{da_{2}}{dt} = \frac{-a_{2}\,^{2} + a_{1}\,^{2} + b_{3}\,^{2}}{4a_{1}b_{3}} + \frac{-a_{2}\,^{2} + a_{3}\,^{2} + b_{1}\,^{2}}{4a_{3}b_{1}} - \frac{1}{2}a_{2}H_{126} - \frac{1}{2}a_{2}H_{234} \,, \label{ODE2}
\end{equation}
\begin{equation}
\frac{da_{3}}{dt} = \frac{- a_{3}\,^{2} + a_{2}\,^{2} + b_{1}\,^{2}}{4a_{2}b_{1}} + \frac{-a_{3}\,^{2} + a_{1}\,^{2} + b_{2}\,^{2}}{4a_{1}b_{2}} + \frac{1}{2}a_{3}H_{135} - \frac{1}{2}a_{3}H_{234} \,, \label{ODE3}
\end{equation}
\begin{equation}
\frac{db_{1}}{dt} = \frac{- b_{1}\,^{2} + a_{2}\,^{2} + a_{3}\,^{2}}{4a_{2}a_{3}} + \frac{b_{1}\,^{2} - b_{2}\,^{2} - b_{3}\,^{2}}{4b_{2}b_{3}} + \frac{1}{2}b_{1}H_{456} - \frac{1}{2}b_{1}H_{234} \,, \label{ODE4}
\end{equation}
\begin{equation}
\frac{db_{2}}{dt} = \frac{- b_{2}\,^{2} + a_{1}\,^{2} + a_{3}\,^{2}}{4a_{3}a_{1}} + \frac{b_{2}\,^{2} - b_{1}\,^{2} - b_{3}\,^{2}}{4b_{3}b_{1}} + \frac{1}{2}b_{2}H_{456} + \frac{1}{2}b_{2}H_{135} \,, \label{ODE5}
\end{equation}
\begin{equation}
\frac{db_{3}}{dt} = \frac{- b_{3}\,^{2} + a_{1}\,^{2} + a_{2}\,^{2}}{4a_{1}a_{2}} + \frac{b_{3}\,^{2} - b_{1}\,^{2} - b_{2}\,^{2}}{4b_{1}b_{2}} - \frac{1}{2}b_{3}H_{126} + \frac{1}{2}b_{3}H_{456} \,, \label{ODE6}
\end{equation}
\begin{equation}
\frac{dH_{126}}{dt} + H_{126}\left(\frac{1}{a_{1}}\frac{da_{1}}{dt} + \frac{1}{a_{2}}\frac{da_{2}}{dt} + \frac{1}{b_{3}}\frac{db_{3}}{dt}\right) = \left(\frac{a_{1}a_{2}}{b_{1}b_{2}} - 1\right)\left(\frac{b_{3}}{a_{1}a_{2}}H_{126} + \frac{a_{3}b_{2}}{a_{1}a_{2}\,^{2}}H_{135}\right) \,, \label{ODE7}
\end{equation}
\begin{equation}
\frac{dH_{135}}{dt} + H_{135}\left(\frac{1}{a_{1}}\frac{da_{1}}{dt} + \frac{1}{a_{3}}\frac{da_{3}}{dt} + \frac{1}{b_{2}}\frac{db_{2}}{dt}\right) = 0 \,. \label{ODE8}
\end{equation}
Thus we have obtained $G_{2}T$ equations (\ref{ODE1})--(\ref{ODE8}) under the cohomogeneity one ansatz with principal orbits $S^{3} \times S^{3}$.
It should be noticed that dilaton is given by (\ref{dPhidt}), which is equivalent to
\begin{equation}
\frac{d\Phi}{dt} = - H_{126} + H_{456} + H_{135} - H_{234} \,. \label{ODE9}
\end{equation}
Note that it is automatically assured by (\ref{alg equations})--(\ref{ODE8}) that the field strength $F$ is closed two-form, $dF = 0$.
\subsection{$G_{2}T$ solutions}
In order to construct regular metrics satisfying (\ref{ODE1})--(\ref{ODE8}), we shall look for solutions with bolt singularities at $t=0$.
There exist two types of bolt singularity corresponding to a collapsing $S^{3}$ or $S^{1}$ at $t=0$.
In the torsion free case, i.e. $H_{\mu\nu\rho} = 0$, these equations were studied by Cveti$\check{\text{c}}$, Gibbons, L$\ddot{\text{u}}$ and Pope \cite{CGLP1}\cite{CGLP2} and they found Ricci-flat metrics with $G_{2}$ holonmy.
These equations are also obtained from $d\Omega = 0$ and $d*\Omega = 0$.
Let us turn to non-zero flux.
Although (\ref{ODE1})--(\ref{ODE8}) admit a Taylor expansion around $t=0$, numerical analysis shows that global solution doesn't exist for general six radial functions $a_{i}(t),b_{i}(t)$.
Hence we study the reduced case, 
\begin{equation}
a_{1}(t) = a_{2}(t) \equiv a(t) \;,\; b_{1}(t) = b_{2}(t) \equiv b(t) \text{.} \label{reduce4}
\end{equation}
Note that the Taylor expansion around $t=0$ gives rise to $H_{135} = 0$.

By (\ref{reduce4}) and $H_{135} = 0$ the reduced system is given by 
\begin{eqnarray}
& & \frac{1}{a}\frac{da}{dt} = \frac{b}{4aa_{3}} + \frac{a_{3}}{4ab} - \frac{a}{4a_{3}b} + \frac{b_{3}}{4a^{2}} -\frac{1}{2}H_{126} \,, \label{RODE1}\\
& & \frac{1}{b}\frac{db}{dt} = - \frac{b}{4aa_{3}} + \frac{a_{3}}{4ab} + \frac{a}{4a_{3}b} - \frac{b_{3}}{4b^{2}} + \frac{a^{2}}{2b^{2}}H_{126} \,, \label{RODE2}\\
& & \frac{1}{a_{3}}\frac{da_{3}}{dt} = \frac{b}{2aa_{3}} - \frac{a_{3}}{2ab} + \frac{a}{2a_{3}b} \,, \label{RODE3}\\
& & \frac{1}{b_{3}}\frac{db_{3}}{dt} = - \frac{b_{3}}{4a^{2}} + \frac{b_{3}}{4b^{2}} + \frac{1}{2}\left(\frac{a^{2}}{b^{2}} - 1\right)H_{126} \,, \label{RODE4}\\
& & \frac{dH_{126}}{dt} = - H_{126}\left[\frac{b^{2} + a_{3}^{2} - a^{2}}{2aa_{3}b} + \frac{b_{3}(5b^{2} - 3a^{2})}{4a^{2}b^{2}}\right] - \frac{1}{2}\left(\frac{a^{2}}{b^{2}} - 3\right)H_{126}^{2} \,, 
\label{RODE5}\\
& & \frac{d\Phi}{dt} = H_{126}\left(\frac{a^{2}}{b^{2}} - 1\right) \,. \label{RODE6}
\end{eqnarray}
Then, the algebraic equations (\ref{alg equations}) are written as
\begin{eqnarray}
H_{456} & = & \frac{a^{2}}{b^{2}}H_{126} \,, \nonumber \\
F_{12} & = & \pm \sqrt{- \frac{1}{2}\left(\frac{b_{3}}{a^{2}}H_{126}\right)} \,, \nonumber \\
F_{45} & = & \frac{a^{2}}{b^{2}}F_{12} \,, \nonumber \\
F_{67} & = & F_{45} - F_{12} \,. \label{reduced alg}
\end{eqnarray}
Now let us solve the equations (\ref{RODE1})--(\ref{RODE6}). 
We put the form 
\begin{equation}
b_{3}(t) = \frac{1}{h(t)}\beta_{3}(t) \,, \label{beta3 to b3}
\end{equation}
where the radial function $\beta_{3}(t)$ is defined as the solution to the ensuing equations, 
\begin{eqnarray}
\frac{1}{a}\frac{da}{dt} & = & \frac{b}{4aa_{3}} + \frac{a_{3}}{4ab} - \frac{a}{4a_{3}b} + \frac{\beta_{3}}{4a^{2}} \,, \label{TFODE1}\\
\frac{1}{b}\frac{db}{dt} & = & - \frac{b}{4aa_{3}} + \frac{a_{3}}{4ab} + \frac{a}{4a_{3}b} - \frac{\beta_{3}}{4b^{2}} \,, \label{TFODE2}\\
\frac{1}{a_{3}}\frac{da_{3}}{dt} & = & \frac{b}{2aa_{3}} - \frac{a_{3}}{2ab} + \frac{a}{2a_{3}b} \,, \label{TFODE3}\\
\frac{1}{\beta_{3}}\frac{d\beta_{3}}{dt} & = & - \frac{\beta_{3}}{4a^{2}} + \frac{\beta_{3}}{4b^{2}} \,, \label{TFODE4}
\end{eqnarray}
which are obtained by setting $H_{126} = 0$ in (\ref{RODE1})--(\ref{RODE6}).
The function $h(t)$ is determined as follows.
From (\ref{RODE1})--(\ref{RODE4}), (\ref{beta3 to b3})--(\ref{TFODE4}), we find the relation 
\begin{equation}
\frac{1}{h}\frac{dh}{dt} = - \frac{d\Phi}{dt} \,, \label{h and dilaton}
\end{equation}
which yields
\begin{equation}
\frac{1}{\beta_{3}}\frac{d\beta_{3}}{dt} = \frac{1}{2(h - 1)}\frac{dh}{dt} \,.
\end{equation}
Thus we have
\begin{equation}
h(t) = k\beta_{3}(t)^{2} + 1 \,. \label{h(t)}
\end{equation}
Then the relation (\ref{beta3 to b3}) is given by
\begin{equation}
b_{3}(t) = \frac{1}{k\beta_{3}(t)^{2} + 1}\beta_{3}(t)  \label{b3}
\end{equation}
and using (\ref{RODE6}), (\ref{h and dilaton}) and (\ref{h(t)}) leads to
\begin{equation}
H_{126} = - \frac{k\beta_{3}\,^{3}}{2a^{2}(k\beta_{3}^{2} + 1)} \label{flux component H126} \,.
\end{equation}
For the positive constant $k$, the algebraic equations (\ref{reduced alg}) give $H_{456}$ and $F_{\mu\nu}$,
\begin{eqnarray}
& & H_{456} = - \frac{k\beta_{3}\,^{3}}{2b^{2}(k\beta_{3}^{2} + 1)} \;,\; F_{12} = \pm \frac{\sqrt{k}\beta_{3}\,^{2}}{2a^{2}(k\beta_{3}\,^{2} + 1)} \,, \nonumber \\
& & F_{45} = \pm \frac{\sqrt{k}\beta_{3}\,^{2}}{2b^{2}(k\beta_{3}\,^{2} + 1)} \;,\; F_{67} = \pm \frac{\sqrt{k}\beta_{3}\,^{2}}{2(k\beta_{3}\,^{2} + 1)}\left(- \frac{1}{a^{2}} + \frac{1}{b^{2}}\right) \,.
\end{eqnarray}
Thus,  under the ansatz (\ref{beta3 to b3}), solutions to (\ref{RODE1})--(\ref{RODE6}) and (\ref{reduced alg}) are written as
\begin{eqnarray}
g & = & dt^{2} + a(t)^{2}\{(\sigma_{1} - \Sigma_{1})^{2} + (\sigma_{2} - \Sigma_{2})^{2}\} + a_{3}(t)^{2}(\sigma_{3} - \Sigma_{3})^{2} \nonumber \\
  &   &  + b(t)^{2}\{(\sigma_{1} + \Sigma_{1})^{2} + (\sigma_{2} + \Sigma_{2})^{2}\} + \frac{1}{(k\beta_{3}(t)\,^{2} + 1)^{2}}\beta_{3}(t)^{2}(\sigma_{3} + \Sigma_{3})^{2} \,, 
  \label{metric}
\end{eqnarray}
\begin{equation}
\Phi(t) = \log{\frac{m}{k\beta_{3}(t)^{2} + 1}} \,, \label{dilaton}
\end{equation}
\begin{equation}
H = - \frac{k\beta_{3}(t)^{3}}{2a(t)^{2}(k\beta_{3}(t)\,^{2} + 1)}e^{126} - \frac{k\beta_{3}(t)^{3}}{2b(t)^{2}(k\beta_{3}(t)\,^{2} + 1)}e^{456} \,, \label{3-form flux}
\end{equation}
\begin{equation}
F = \frac{\sqrt{k}\beta_{3}\,^{2}}{2a^{2}(k\beta_{3}\,^{2} + 1)}e^{12} + \frac{\sqrt{k}\beta_{3}\,^{2}}{2b^{2}(k\beta_{3}\,^{2} + 1)}e^{45} + \frac{\sqrt{k}\beta_{3}\,^{2}}{2(k\beta_{3}\,^{2} + 1)}\left(- \frac{1}{a^{2}} + \frac{1}{b^{2}}\right)e^{67} \,, \label{field strength}
\end{equation}
where $m$ is an arbitrary constant
\footnote{If we set $k = tan^{2}\delta$, $m = \frac{1}{cos^{2}\delta}$ and hence $m=k+1$, then these solutions are coincident with the solutions obtained by $\text{r}^{-1}$Tr transformation in the paper \cite{MS}. 
When we take a limit $k \rightarrow \infty$,~$m \rightarrow \infty$ keeping $\frac{k}{m} \rightarrow 1$, then $\delta$ is equal to $\frac{\pi}{2}$, which corresponds to T-dual transformation.}.
This formula includes additional one parameter $m$ comparing to the formula obtained by $\mathrm{r}^{-1}\mathrm{T}\mathrm{r}$ transformation\cite{MS}.
Now the problem of finding the solutions to $G_{2}T$ equations reduced to that of finding the solutions to the torsion free equations(\ref{TFODE1})--(\ref{TFODE4}).
It should be noticed that the torsion free equations were investigated in the study of Ricci-flat metrics with $G_{2}$ holonomy \cite{CGLP1}.
It was proved\cite{Bazaikin1} that there exist the solutions with a collapsing $S^{1}$ for the torsion free equations (\ref{TFODE1})--(\ref{TFODE4}).

A regular solution to (\ref{TFODE1})--(\ref{TFODE4}) representing a Ricci-flat metric with $G_{2}$ holonomy is known \cite{BGGG},
\begin{eqnarray}
a_{3}(r) & = & - \frac{1}{2}r \;,\; a(r) = \frac{1}{4}\sqrt{3(r - l)(r + 3l)} \,, \nonumber \\
\beta_{3}(r) & = & l\sqrt{\frac{r^{2} - 9l^{2}}{r^{2} - l^{2}}} \;,\; b(r) = - \frac{1}{4}\sqrt{3(r + l)(r - 3l)} \,,
\end{eqnarray}
where $r$ is defined by $dt = \frac{3}{2}\frac{l}{\beta_{3}}dr$ and $l$ is a scaling parameter $(l>0)$.
The corresponding solutions of $G_{2}T$ equations are given by
\begin{eqnarray}
g & = & \frac{9}{4}\frac{r^{2} - l^{2}}{r^{2} - 9l^{2}}dr^{2} + \frac{3}{16}(r - l)(r + 3l)\{(\sigma_{1} - \Sigma_{1})^{2} + (\sigma_{2} - \Sigma_{2})^{2}\} + \frac{1}{4}r^{2}(\sigma_{3} - \Sigma_{3})^{2} \nonumber \\
  &   &  + \frac{3}{16}(r + l)(r - 3l)\{(\sigma_{1} + \Sigma_{1})^{2} + (\sigma_{2} + \Sigma_{2})^{2}\} + \frac{l^{2}(r^{2} - 9l^{2})}{kl^{2}(r^{2} - 9l^{2}) + r^{2} - l^{2}}(\sigma_{3} + \Sigma_{3})^{2} \,, \nonumber \\
H & = & - \frac{8k}{3\left(1 + k\frac{r^{2} - 9l^{2}}{r^{2} - l^{2}}\right)}\sqrt{\frac{r^{2} - 9l^{2}}{r^{2} - l^{2}}}\left[\frac{r - 3l}{3(r - l)^{2}(r + l)}e^{126} + \frac{r + 3l}{(r + l)^{2}(r - l)}e^{456}\right] \,, \nonumber \\
F  & = & - \frac{8\sqrt{k}}{3\left(1 + k\frac{r^{2} - 9l^{2}}{r^{2} - l^{2}}\right)}\left[\frac{r - 3l}{3(r - l)^{2}(r + l)}e^{12} + \frac{r + 3l}{(r + l)^{2}(r - l)}e^{45} + \frac{2(r^{2} - 3l^{2})}{(r^{2} - l^{2})^{2}}e^{67}\right] \,, \nonumber \\
\Phi & = & -m\log\left[1 + k\frac{r^{2} - 9l^{2}}{r^{2} - l^{2}}\right] \,. \label{exact sol.T}
\end{eqnarray}
These solutions include two free parameters $l$ and $k$
\footnote{For $m = k + 1$, the equation (\ref{exact sol.T}) is also obtained by solution-generating method \cite{MS}.}.
When $k=0$, the solution reduces to the Ricci-flat metric with $G_{2}$ holonomy.
The parameter $r$ is lager than $3l$ by a regular condition and the other parameter $k$ represents deformation from the Ricci-flat metric.

The metric (\ref{exact sol.T}) is an ALC(Asymptotically Locally Conical) metric and the Riemannian tensor is non-singular in the region $3l\leq r<\infty$.
Further the scalar curvature $R$, $F^{2} = \frac{1}{2}F_{\mu\nu}F^{\mu\nu}$ and $H^{2} = \frac{1}{6}H_{\mu\nu\rho}H^{\mu\nu\rho}$ are finite but their integrations diverge.
The cause of the divergence comes from higher powers of $r$ in the volume form, $\sqrt{g} \sim r^{5}$.
\subsubsection{$S^{3}$-bolt solution}
From now on, we explain the reason why we can put the relation (\ref{beta3 to b3}).
The boundary condition representing a collapsing $S^{3}$ at $t=0$ imposes $b(t),b_{3}(t) \rightarrow 0$ for $t \rightarrow 0$.
Indeed, the series solution around $t=0$ takes the following form, 
\begin{eqnarray}
a(t) & = & p_{1} + \frac{1}{16p_{1}^{2}}t^{2} - \frac{7 - 64p_{1}^{2}p_{2}}{2560p_{1}^{3}}t^{4} + \cdots \,, \nonumber \\
a_{3}(t) & = & - p_{1} -  \frac{1}{16p_{1}^{2}}t^{2} + \frac{6 + 128p_{1}^{2}p_{2}}{2560p_{1}^{3}}t^{4} + \cdots \,, \nonumber \\
b(t) & = & - \frac{1}{4}t + p_{2}t^{3} - \frac{1 + 1344p_{1}^{2}p_{2} - 98304p_{1}^{4}p_{2}^{2}}{10240p_{1}^{4}}t^{5} + \cdots \,, \nonumber \\
b_{3}(t) & = & \frac{1}{4}t + \frac{128p_{1}^{2}p_{2} + 128p_{1}^{4}p_{3} - 1}{64p_{1}^{2}}t^{3} \nonumber \\
         &   & - \frac{216p_{1}^{2}p_{2} - 16896p_{1}^{4}p_{2}^{2} + 240p_{1}^{4}p_{3} - 30720p_{1}^{6}p_{2}p_{3} - 10240p_{1}^{8}p_{3}^{2} - 1}{640p_{1}^{4}}t^{5} + \cdots \,, \nonumber \\
H_{126}(t) & = & p_{3}t^{3} + \left(8p_{1}^{2}p_{3}^{2} + 24p_{2}p_{3} - \frac{5p_{3}}{16p_{1}^{2}}\right)t^{5} + \cdots \,, \label{S3bolt expansion}
\end{eqnarray}
where $p_{1}$, $p_{2}$ and $p_{3}$ are free parameters.
Then the metric behaves as 
\begin{eqnarray}
g & \rightarrow & dt^{2} + p_{1}^{2}[ (\sigma_{1} - \Sigma_{1})^{2} + (\sigma_{2} - \Sigma_{2})^{2} + (\sigma_{3} - \Sigma_{3})^{2} ] \nonumber \\
  &             &  + \frac{t^{2}}{16}[ (\sigma_{1} + \Sigma_{1})^{2} + (\sigma_{2} + \Sigma_{2})^{2} + (\sigma_{3} + \Sigma_{3})^{2} ] \,.
\end{eqnarray}
Thus the metric has a removable bolt singularity ($S^{3}$-bolt solution).
The parameter $p_{1}$ is a scaling parameter and $p_{1} = \frac{3}{2}l$ for the solutions (\ref{exact sol.T}), while $p_{2}$ parametrizes a family of the regular solutions to (\ref{TFODE1})--(\ref{TFODE4}).
The remaining parameter $p_{3}$ is essentially the deformation parameter $k$ associated with 3-form flux (\ref{flux component H126}), 
\begin{equation}
p_{3} = - \frac{k}{128p_{1}^{2}} \,.
\end{equation}
Actually when $p_{3} = 0$, the flux component $H_{126}(t)$ vanishes and the corresponding solution reduces to Ricci-flat metrics with a collapsing $S^{3}$ at $t=0$ \cite{CGLP1}.
The radial functions $a(t)$, $b(t)$ and $a_{3}(t)$ are independent on $p_{3}$, which means that these functions are the same as those of the Ricci-flat metrics.
On the other hand, $b_{3}(t)$ depends on $p_{3}$ so that this is different from $\beta_{3}(t)$ satisfying (\ref{TFODE1})--(\ref{TFODE4}).
Namely $b_{3}(t)$ is only deformed by the 3-form flux and hence we can take the form (\ref{beta3 to b3}).

It should be noticed that we can obtain all $S^{3}$-bolt solutions of (\ref{RODE1})--(\ref{RODE6}) from regular solutions of (\ref{TFODE1})--(\ref{TFODE4}).
In particular, the parameters $p_{1} = \frac{3}{2}l$ and $p_{2} = - \frac{1}{768p_{1}^{2}}$ correspond to the solutions (\ref{exact sol.T}).
The numerical calculation requires the inequality of the parameter $p_{2}$ for each values of deformation parameter $p_{3}$ in order to extend the solution in large $t$ region.
For example when $p_{1}=1$ and $p_{3} = -900$, the parameter $p_{2}$ is restricted by the inequality $-7.029 \times 10^{3} \leq p_{2} \leq 1.321 \times 10^{-3}$ (see Figure 1).
\\
\begin{figure}[!h]
\begin{center}
\includegraphics[width=8cm]{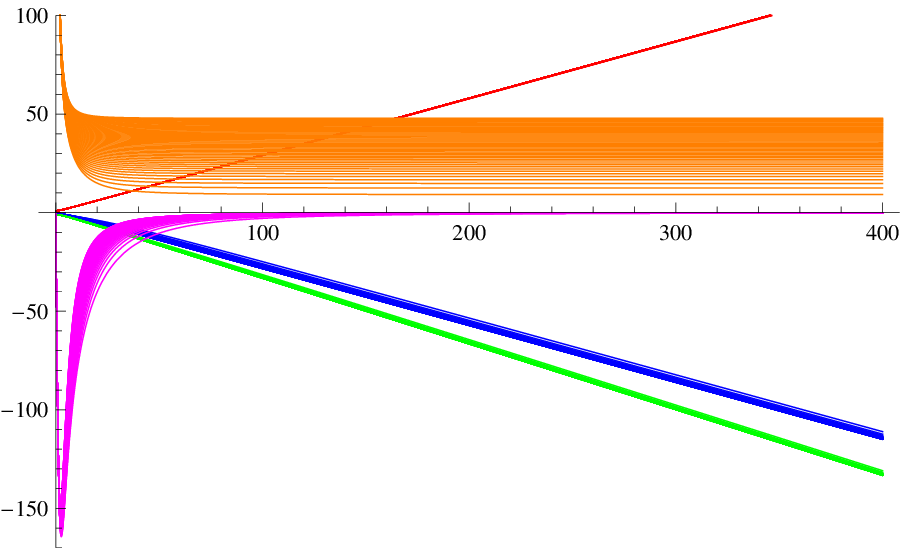}
\includegraphics[width=8cm]{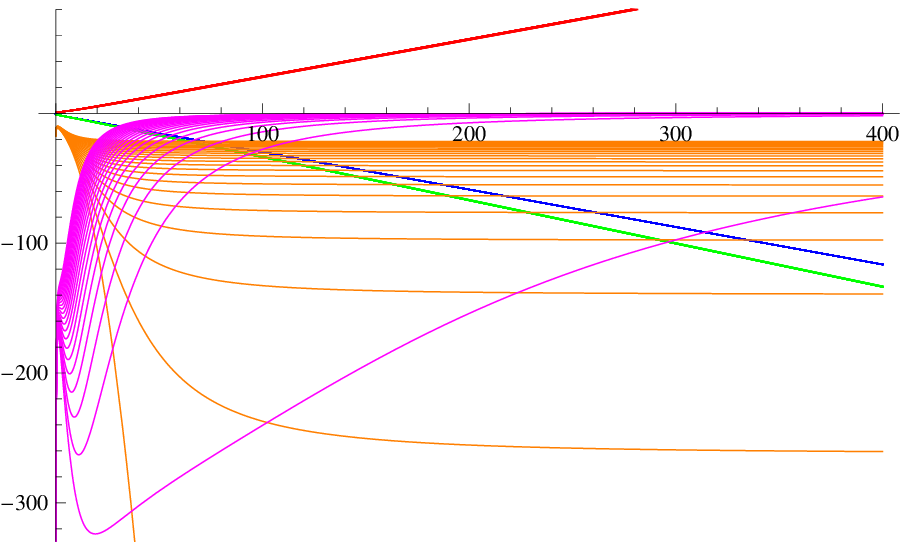}
\caption{The both figures show $S^{3}$-bolt solutions.
The left figure is the solution with $p_{1} = 1$, $-1.321 \times 10^{-3} \leq p_{2} \leq 1.321 \times 10^{-3}$ and $p_{3}=-900$
and the right figure is the solution with $p_{1} = 1$, $- 7.029 \times 10^{3} \leq p_{2} \leq - 6.852 \times 10^{3}$ and $p_{3}=-900$. 
The red, green, blue, orange and magenta lines represent components $a(t)$, $a_{3}(t)$, $b(t)$, $b_{3}$ and $H_{126}(t)$, respectively.}
\end{center}
\end{figure}
\\
\subsubsection{$T^{1,1}$-bolt solution}
Regular solutions with a collapsing $S^{1}$ takes the following form around $t=0$,
\begin{eqnarray}
a(t) & = & p_{1} + \frac{p_{2} - 2p_{1}^{2}p_{3}}{8p_{1}}t - \frac{3p_{2}^{2} - 16p_{1}^{2} -12p_{1}^{2}p_{2}p_{3} + 12p_{1}^{4}p_{3}^{2}}{128p_{1}^{3}}t^{2} + \cdots \,, \nonumber \\ 
a_{3}(t) & = & t - \frac{16p_{1}^{2} - p_{2}^{2} + 4p_{1}^{2}p_{2}p_{3} - 4p_{1}^{4}p_{3}^{2}}{96p_{1}^{4}}t^{3} + \cdots \,, \nonumber \\
b(t) & = & p_{1} - \frac{p_{2} - 2p_{1}^{2}p_{3}}{8p_{1}}t - \frac{3p_{2}^{2} - 16p_{1}^{2} -12p_{1}^{2}p_{2}p_{3} + 12p_{1}^{4}p_{3}^{2}}{128p_{1}^{3}}t^{2} + \cdots \,, \nonumber \\ 
b_{3}(t) & = & p_{2} + \frac{4p_{2}^{3} - 16p_{2}p_{3}^{2}}{64p_{1}^{4}}t^{2} + \cdots \,, \nonumber \\
H_{126}(t) & = & p_{3} + \frac{2p_{1}^{2}p_{3}^{2} - p_{2}p_{3}}{4p_{1}^{2}}t - \frac{-9p_{2}^{2}p_{3} + 8p_{1}^{2}p_{3} + 28p_{1}^{2}p_{2}p_{3}^{2} - 20p_{1}^{4}p_{3}^{3}}{32p_{1}^{4}}t^{2} + \cdots \,, 
\label{S1bolt expansion}
\end{eqnarray}
where $p_{1}$, $p_{2}$ and $p_{3}$ are free parameters.
Then the metric behaves as 
\begin{eqnarray}
g & \longrightarrow  & dt^{2} + t^{2}(\sigma_{3} - \Sigma_{3})^{2} + p_{1}^{2}( \sigma_{1}^{2} + \Sigma_{1}^{2} + \sigma_{2}^{2} + \Sigma_{2}^{2} ) \nonumber \\
  &   &  + p_{2}^{2}(\sigma_{3} + \Sigma_{3})^{2} \,. 
\end{eqnarray}
When this metric has $T^{1,1}$-bolt, $p_{1}$ and $p_{2}$ have relation (\cite{CGLP2})
\begin{equation}
p_{2} = \pm \sqrt{\frac{2}{3}}p_{1} \,.
\end{equation}
In the series (\ref{S1bolt expansion}), all components are deformed by the 3-form flux, which implies that the relation like (\ref{beta3 to b3}) doesn't exist in the solution.
A numerical analysis indicates that the series solution (\ref{S1bolt expansion}) is extended to large $t$ region (see Figure 2)
\footnote{In the Ricci flat case, the existence of $G_{2}$ metrics with an collapsing $S^{1}$ was proved in \cite{Bazaikin1}.}.
\begin{figure}[!h]
\begin{center}
\includegraphics[width=8cm]{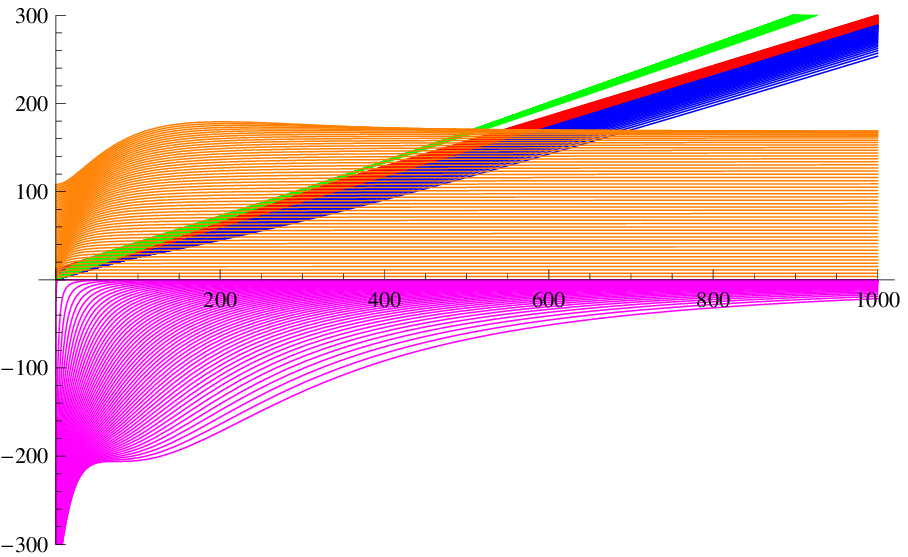}
\includegraphics[width=8cm]{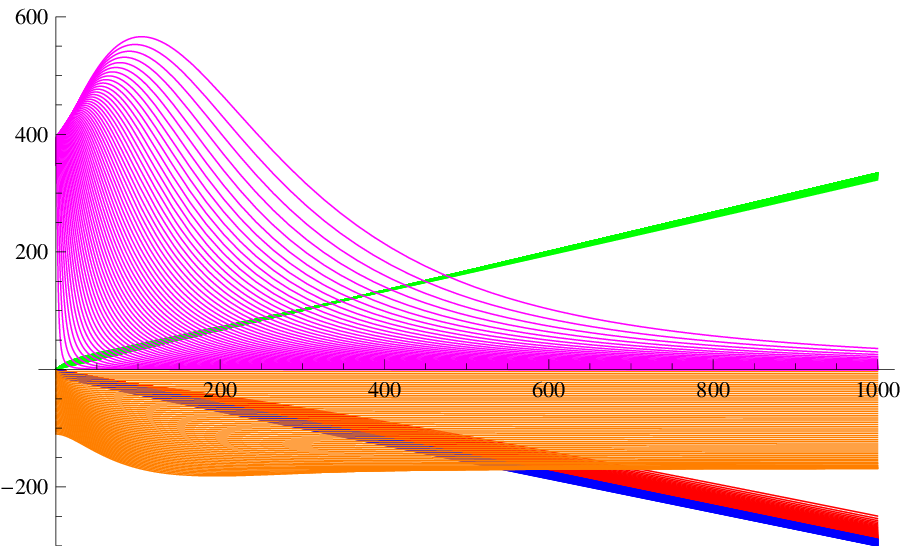}
\caption{The left figure shows a $T^{1,1}$-bolt solution with $p_{1} = \sqrt{\frac{3}{2}}p_{2}$, $0< p_{2} < 12$ and $p_{3} = - \frac{1}{288}$. 
The right figure shows a $T^{1,1}$-bolt solution with $p_{1} = \sqrt{\frac{3}{2}}p_{2}$, $-12 < p_{2} <0$ and $p_{3} = \frac{1}{288}$.
The red, green, blue, orange and magenta lines represent components $a(t)$, $a_{3}(t)$, $b(t)$, $b_{3}$ and $H_{126}(t)$, respectively.}
\end{center}
\end{figure}
\section{$Spin(7)$ solutions in 8-dimensional Abelian heterotic supergravity}
\subsection{$Spin(7)$-structure}
A $Spin(7)$-structure over $M_8$ is a principal subbundle with fiber $Spin(7)$ of the frame bundle $\mathcal{F}(M_8)$ \cite{Joyce}. There is a one to one correspondence between $Spin(7)$ structures and $Spin(7)$ invariant
4-forms $\Psi$. The standard form of $\Psi$ is given by
\begin{eqnarray}\label{spin}
\Psi&=&e^{8123}+e^{8145}+e^{8167}+e^{8264}+e^{8257}+e^{8347}+e^{8356} \nonumber\\
& &-e^{4567}-e^{2367}-e^{2345}-e^{3157}-e^{1346}-e^{1256}-e^{1247},
\end{eqnarray}
where $\{e^\mu, \mu=1, \cdots,8 \}$ is an orthonormal frame of the metric $g=\sum_{\mu=1}^8 e^\mu \otimes e^\mu$. The 4-form $\Psi$ is self-dual $*\Psi=\Psi$ for a volume form $vol=e^{12\cdots 8}$.
The Lee form is defined by
\begin{equation}
\Theta=-\frac{1}{7}*(*d\Psi \wedge \Psi).
\end{equation}
Then, there always exists a unique connection $\nabla^T$ preserving the $Spin(7)$ structure, $\nabla^T \Psi=\nabla^T g=0$, with 3-form torsion \cite{I2}
\begin{equation}
T=*d\Psi-\frac{7}{6}*(\Theta \wedge \Psi).
\end{equation}
This is different from the case of $G_2$, where we required an additional condition (\ref{G2defeq1}).
However, except for this point we have very similar supersymmetry equations associated with Abelian heterotic supergravity. The Lee form is an exact 1-form, $\Theta=(6/7)d\Phi$, and hence after identification $T=H$ the equations are written as
\begin{eqnarray}
& &H =  e^{\Phi}*d(e^{-\Phi}\Psi) \label{Spin1} \,,\\
& &d\Phi= - \frac{1}{6}*(*d\Psi \wedge \Psi) \label{Spin2} \,, \\
& &*(\Psi \wedge F)= F \label{Spin3} \,, \\ 
& &dH  =  F \wedge F \,, \label{Spin4}
\end{eqnarray}
which correspond to $G_2T$ equations.
As with $G_2$, we call these equations $Spin(7)$ with torsion equations.
\subsection{3-Sasakian ansatz}
 In order to construct explicit solutions of $Spin(7)$ with torsion equations, we will consider a special case that gives rise to ordinary differential equations and generalizes many known examples.
Let us assume the following geometrical condition for 7-dimensional manifolds. Let $\pi: P \rightarrow B$ be a principal SO(3)-bundle over a compact self-dual Einstein manifold (or orbifold) $B$.
Then, it is known that the total space $P$ admits a 3-Sasakian structure.
For details on 3-Sasakian structure we refer the reader to \cite{SurveysDG}.
The connection 1-forms $\phi^i~(i=1,2,3)$ will be chosen so that the corresponding curvature 2-forms
\begin{equation}
\omega^i=d\phi^i+\frac{1}{2}\sum_{j,k=1}^3 \epsilon_{ijk} \phi^j \wedge \phi^k
\end{equation}
are self-dual 2-forms on $B$,
\begin{equation}
\omega^1=\theta^{45}+\theta^{67},~~\omega^2=\theta^{64}+\theta^{57},~~\omega^3=\theta^{47}+\theta^{56},
\end{equation}
which implies
\begin{eqnarray}
& &\omega^1 \wedge \omega^2=\omega^1 \wedge \omega^3=\omega^2 \wedge \omega^3=0~,\\
& &\omega^1 \wedge \omega^1=\omega^2 \wedge \omega^2=\omega^3 \wedge \omega^3=2 vol_B~,\\
& &vol_B=\theta^{4567}~~~~(\mbox{volume form on $B$})
\end{eqnarray}
and
\begin{eqnarray}
& &d\omega^1=\omega^2 \wedge \phi^3-\phi^2 \wedge \omega^3\,, \nonumber\\
& &d\omega^2=\omega^3 \wedge \phi^1-\phi^3 \wedge \omega^1\,, \nonumber\\
& &d\omega^3=\omega^1 \wedge \phi^2-\phi^1 \wedge \omega^2\,,
\end{eqnarray}
 On $P$ the metric
\begin{equation}
g_S=\sum_{i=1}^3 \phi^i \otimes \phi^i+\sum_{a=4}^7 \theta^a \otimes \theta^a
\end{equation}
is a 3-Sasakian metric. Indeed, the Killing vector fields $\xi_i~(i=1,2,3)$ dual to the 1-forms $\phi^i$ give the characteristic vector fields of the 3- Sasakian structure satisfying
the relations $[\xi_i,\, \xi_j]=\epsilon_{ijk}\, \xi_k$.

Now we consider a $Spin(7)$ structure on an 8-dimensional manifold 
 $ M=\textrm{\boldmath $R$}_{+} \times P$. Using the 3-Sasakian structure we consider the following metric with four radial functions
 $\{ a_i(t)\}=\{ a(t),\,b(t),\,c(t)\}$ and $f(t)$:
 \begin{equation}
 g=dt^2+\sum_{i=1}a_i(t)^2 \phi^i \otimes \phi^i+f(t)^2 \sum_{a=4}^7 \theta^a \otimes \theta^a.
\end{equation}
For this metric an orthonormal frame $e^\mu $ $(\mu=1,\cdots,8)$ is introduced by
\begin{equation}
e^i= a_i(t) \phi^i,~~e^a=f(t) \theta^a,~~e^8=dt.
\end{equation}
Then the 1-forms $e^\mu$ lead to a $Spin(7)$ structure $\Psi$ on $ M=\textrm{\boldmath $R$}_{+} \times P$ given by (\ref{spin}). In terms of $\phi^i$, $\omega^i$ and $vol_B$ 
we have
\begin{eqnarray}
\Psi&=&a b c\, dt \wedge \phi^1 \wedge \phi^2 \wedge \phi^3-f^4 vol_B+f^2 (a dt \wedge \phi^1-b c \phi^2 \wedge \phi^3)\wedge \omega^1 \nonumber\\
& &+f^2 (b dt \wedge \phi^2-a c \phi^3 \wedge \phi^1)\wedge \omega^2+f^2 (c dt \wedge \phi^3-a b \phi^1 \wedge \phi^2)\wedge \omega^3.
\end{eqnarray}
A straightforward computation gives
\begin{equation}
d\Psi=\Psi_0 dt \wedge vol_B+\Psi_1 dt \wedge \phi^2 \wedge \phi^3 \wedge \omega^1+\Psi_2 dt \wedge \phi^3 \wedge \phi^1 \wedge \omega^2+\Psi_3 dt \wedge \phi^1 \wedge \phi^2 \wedge \omega^3,
\end{equation}
where
\begin{eqnarray}
& &\Psi_0=-4 f^3 \frac{df}{dt}-2 f^2(a+b+c), \nonumber\\
& &\Psi_1=-abc+f^2 a-f^2b-f^2 c-2 b c f\frac{df}{dt}-f^2 \frac{d(bc)}{dt}, \nonumber\\
& &\Psi_2=-abc-f^2 a+f^2b-f^2 c-2 a c f\frac{df}{dt}-f^2 \frac{d(ac)}{dt}, \nonumber\\
& &\Psi_3=-abc-f^2 a-f^2b+f^2 c-2 a b f\frac{df}{dt}-f^2 \frac{d(ab)}{dt}.
\end{eqnarray}
The Ricci-flat metrics with $Spin(7)$ holonomy satisfy the condition $d\Psi=0$. In our case this is explicitly given by
the following first-order differential equations \cite{CGLP1}:
\begin{eqnarray}\label{Ricci}
& &\frac{da}{dt}=-\frac{a^2-(b-c)^2}{2 bc}+\frac{a^2}{f^2},\,~~ \nonumber\\
& &\frac{db}{dt}=-\frac{b^2-(c-a)^2}{2 ac}+\frac{b^2}{f^2},\,~~ \nonumber\\
& &\frac{dc}{dt}=-\frac{c^2-(a-b)^2}{2 ab}+\frac{c^2}{f^2},\,~~ \nonumber\\
& &\frac{df}{dt}=-\frac{a+b+c}{2f}\,.
\end{eqnarray}
 Now we turn to the $Spin(7)$ with torsion equations (\ref{Spin1})--(\ref{Spin4}). The dilaton $\Phi$ is calculated as
\begin{equation}\label{di}
\frac{d\Phi}{dt}=-\frac{1}{6 f^2} \left( \frac{\Psi_0}{f^2}+\frac{2 \Psi_1}{bc}+ \frac{2 \Psi_2}{ac}+ \frac{2 \Psi_3}{ab} \right)
\end{equation}
and th 3-form flux is given by
\begin{equation}\label{torsion}
H=H_{\phi} \phi^1 \wedge \phi^2 \wedge \phi^3+\sum_{i=1}^3 H_{i}\phi^i \wedge \omega^i,
\end{equation}
where
\begin{eqnarray}
& &H_{\phi}=-a b c\left( \frac{\Psi_0}{f^4}+\frac{d\Phi}{dt} \right)\,, ~~~
H_{1}=-a f^2\left( \frac{\Psi_1}{b c f^2}+\frac{d\Phi}{dt} \right)\,, \nonumber\\
& &H_{2}=-b f^2\left( \frac{\Psi_2}{a c f^2}+\frac{d\Phi}{dt} \right)\,, ~~~
H_{3}=-c f^2\left( \frac{\Psi_3}{a b f^2}+\frac{d\Phi}{dt} \right)\,.
\end{eqnarray}
If the field strength $F$ takes the form,
\begin{equation}
F=\sum_{i=1}^3F_{ti} dt \wedge d\phi^i+\sum_{i=1}^3 F_i \omega^i+\frac{1}{2}\sum_{i,j=1}^3 F_{ij} \phi^i \wedge \phi^j\,,
\end{equation}
then the generalized self-dual equation (\ref{Spin3}) yields
\begin{equation}
\frac{2 F_1}{f^2}=-\frac{F_{23}}{bc}+\frac{F_{t1}}{a},~~~~\frac{2 F_2}{f^2}=-\frac{F_{31}}{ac}+\frac{F_{t2}}{b},~~~~\frac{2 F_3}{f^2}=-\frac{F_{12}}{ab}+\frac{F_{t3}}{c}.
\end{equation}
From (\ref{torsion}) we obtain that
\begin{eqnarray}
dH&=&\frac{dH_\phi}{dt} dt \wedge \phi^1 \wedge \phi^2 \wedge \phi^3+\sum_{i=1}^3 \frac{dH_i}{dt}dt \wedge \phi^i \wedge \omega^i+2 \sum_{i=1}^3 H_i\, vol_B \nonumber\\
&+& (H_\phi-H_1+H_2+H_3)\phi^2 \wedge \phi^3 \wedge \omega^1+(H_\phi+H_1-H_2+H_3)\phi^3 \wedge \phi^1 \wedge \omega^2 \nonumber\\
&+&
(H_\phi+H_1+H_2-H_3)\phi^1 \wedge \phi^2 \wedge \omega^3\,,
\end{eqnarray}
Using (\ref{Spin4}) and (\ref{di}) we find that the solutions are classified into three types:
\begin{eqnarray}
\mbox{(a)}& &F=d(F_1 \phi^1)\,,~~~
H=F_1 \phi^1 \wedge F\,, \label{case1}\nonumber\\
& &\frac{dF_1}{dt}=a F_1 \left( \frac{2}{f^2}-\frac{1}{bc} \right)\,,~~~\frac{d\Phi}{dt}=-\frac{F_1^2}{a} \left( \frac{2}{f^2}-\frac{1}{bc} \right)\,.\\
\mbox{(b)}& &F=d(F_2 \phi^2)\,,~~~
H=F_2 \phi^2 \wedge F\,, \label{torsion1} \nonumber\\
& &\frac{dF_2}{dt}=b F_2 \left( \frac{2}{f^2}-\frac{1}{ac} \right)\,,~~~\frac{d\Phi}{dt}=-\frac{F_2^2}{b} \left( \frac{2}{f^2}-\frac{1}{ac} \right)\,.\\
\mbox{(c)}& &F=d(F_3 \phi^3)\,,~~~
H=F_3 \phi^3 \wedge F\,, \nonumber\\
& &\frac{dF_3}{dt}=c F_3 \left( \frac{2}{f^2}-\frac{1}{ab} \right)\,,~~~\frac{d\Phi}{dt}=-\frac{F_3^2}{c} \left( \frac{2}{f^2}-\frac{1}{ab} \right)\,.
\end{eqnarray}
Then, for case (a), the $Spin(7)$ with torsion equation (\ref{Spin1}) reduces to the following differential equations:
\begin{eqnarray}\label{torsion2}
& &\frac{da}{dt}=-\frac{a^2-(b-c)^2-F_1^2}{2 bc}+\frac{a^2-F_1^2}{f^2}\,~~, \nonumber\\
& &\frac{db}{dt}=-\frac{b^2-(c-a)^2-F_1^2}{2 ac}+\frac{b^2}{f^2}\,~~, \nonumber\\
& &\frac{dc}{dt}=-\frac{c^2-(a-b)^2-F_1^2}{2 ab}+\frac{c^2}{f^2}\,~~, \nonumber\\
& &\frac{df}{dt}=-\frac{a+b+c}{2f}-\frac{F_1^2}{2 a f}\,~~,
\end{eqnarray}
and (b) and (c) are given by cyclic permutations of $a,b,c$.

Finally, we briefly discuss the solutions to (\ref{Ricci}) and the $Spin(7)$ with torsion equations given by (\ref{torsion2}).
A more detail about the solutions will be reported elsewhere.
The regular condition to the solutions requires
the following boundary conditions at $t=0$:
\begin{eqnarray}
& &\mbox{(I)}~~ a(0)=b(0)=c(0)=0,~~~f(0) \ne 0,\nonumber\\
& &~~~~~~~|a'(0)|=|b'(0)|=|c'(0)|=\frac{1}{2},~~~f'(0)=0, \\
& &\mbox{(II)}~~a(0)=0,~~~b(0)=-c(0),~~~f(0) \ne 0,\nonumber\\
& &~~~~~~~|a'(0)|=2,~~~b'(0)=c'(0),~~~f'(0)=0.
\end{eqnarray}
The equation (\ref{Ricci}) gives Ricci-flat $Spin(7)$ holonomy metrics.
Some explicit solutions satisfying the boundary conditions (I)(II) are constructed in \cite{CGLP1}\cite{CGLP2}\cite{CGLP3}\cite{CGLP4} with the help of numerical calculations, and further the existence of the regular solutions was analytically proved \cite{Bazaikin2}.
The $Spin(7)$ with torsion case (\ref{torsion2}) is slightly different from  the Ricci-flat case.  We find that case (II) admits no solution with non-zero $F_1$, while
 case (I) admits the following series solution around $t=0$:
\begin{eqnarray}\label{SR}
& &a(t)=-\frac{t}{2}+a_3 t^3+a_5 t^5+\cdots~, \nonumber\\
& &b(t)=-\frac{t}{2}+b_3 t^3+b_5 t^5+\cdots~, \nonumber\\
& &c(t)=-\frac{t}{2}+c_3 t^3+c_5 t^5+\cdots~, \nonumber\\
& &f(t)=f_0+\frac{3}{8 f_0}t^2+f_4 t^4+\cdots~, \nonumber\\
& &F_1(t)=h_2 t^2+h_4 t^4+\cdots~.
\end{eqnarray}
Here, the series have four independent  parameters,  $\{a_3, b_3, c_3, f_0, h_2 \}$ with one constraint $a_3+b_3+c_3=2 h_2^2+1/(4 f_0^2)$, and higher coefficients are determined by these parameters.
The series (\ref{SR}) can be extended to an ALC solution of (\ref{case1}) and (\ref{torsion2}) when the parameters are restricted to $b_3=c_3$, and hence $b(t)=c(t)$ is satisfied for all $t$.  
Specifically, the explicit ALC solution with two parameters $ \ell, k$ is given by
\begin{eqnarray}
& &a(r)=-\frac{\ell(r-\ell)\sqrt{(r-3 \ell)(r+\ell)}}{(1+k^2 \ell^2)(r-\ell)^2-4 k^2 \ell^4}, \nonumber\\
& &b(r)=c(r)=-\frac{1}{2}\sqrt{(r-3 \ell)(r+\ell)}, \nonumber\\
& &f(r)=\sqrt{\frac{r^2-\ell^2}{2}}
\end{eqnarray}
together with
\begin{equation}
F_1(r)=\frac{k \ell^2(r-3\ell)(r+\ell)}{(1+k^2 \ell^2)(r-\ell)^2-4 k^2 \ell^4}. 
\end{equation}
Here, we used a radial coordinate $r~(r \ge 3 \ell)$ defined by $dt=(r-\ell)dr/\sqrt{(r-3\ell)(r+\ell)}$.
\section{Conclusion}
We have derived $G_{2}T$ equations in Abelian heterotic supergravity theory. When a $G_{2}$ manifold is locally given by $\textrm{\boldmath $R$}_{+} \times S^{3} \times S^{3}$,
the $G_{2}T$ equations are reduced to ordinary differential equations(\ref{ODE1})--(\ref{ODE8}).
A numerical analysis of these equations shows that a global solution doesn't exist for the general six radial functions $a_{i}(t),b_{i}(t) \, (i=1,\cdots,3)$ 
and thus we study the reduced case, $a_{1}(t) = a_{2}(t)$ and $b_{1}(t) = b_{2}(t)$. 
The Abelian heterotic solutions $(g,H,\varphi,F)$ are obtained from (\ref{metric})--(\ref{field strength}) by using Ricci-flat $G_2$ holonomy metrics.
To construct regular solutions of the reduced $G_{2}T$ equations, 
we have investigated $S^{3}$-bolt solutions and $T^{1,1}$-bolt solutions by numerical analysis.
These solutions are shown graphically in figures 1 and 2.
The formulas (\ref{metric})--(\ref{field strength}) generate only $S^{3}$-bolt solutions.
A problem of finding analytic expressions for $T^{1,1}$-bolt solutions remains as a future work.
We have also derived $Spin(7)$ with torsion equations based on 3-Sasakian manifolds.
The explicit ALC solution to Abelian heterotic supergravity theory has been obtained from these equations.
It is of great interest to study general solutions, which would  shed new light on $G_{2}$ or $Spin(7)$ with torsion geometry.

\end{document}